\documentclass{PoS}

\title{Probability distribution functions in the finite density lattice QCD}

\ShortTitle{Probability distribution functions in the finite density lattice QCD}

\author{\speaker{S.~Ejiri},
Y.~Nakagawa\\
Graduate School of Science and Technology, Niigata University, Niigata 950-2181, Japan\\
        E-mail: \email{ejiri@muse.sc.niigata-u.ac.jp}
}
\author{S.~Aoki, K.~Kanaya, H.~Saito\\
Graduate School of Pure and Applied Sciences, University of Tsukuba, Tsukuba, Ibaraki 305-8571, Japan
}
\author{T.~Hatsuda\\
Theoretical Research Division, Nishina Center, RIKEN, Wako 351-0198, Japan
}
\author{H.~Ohno\\
Fakult\"at f\"ur Physik, Universit\"at Bielefeld, D-33615 Bielefeld, Germany
}
\author{T.~Umeda\\
Graduate School of Education, Hiroshima University, Hiroshima 739-8524, Japan
}
\author{(WHOT-QCD collaboration)}

\abstract{
We study the phase structure of QCD at high temperature and density by lattice QCD simulations adopting a histogram method. 
We try to solve the problems which arise in the numerical study of the finite density QCD, 
focusing on the probability distribution function (histogram). 
As a first step, we investigate the quark mass dependence and the chemical potential dependence of the probability distribution function as a function of the Polyakov loop when all quark masses are sufficiently large, and study the properties of the distribution function. 
The effect from the complex phase of the quark determinant is estimated explicitly. 
The shape of the distribution function changes with the quark mass and the chemical potential. 
Through the shape of the distribution, the critical surface which separates the first order transition and crossover regions in the heavy quark region is determined for the 2+1-flavor case. 
}

\FullConference{The 30th International Symposium on Lattice Field Theory\\
		 June 24 - 29,  2012\\
		 Cairns, Australia}

\begin{document}

\section{Histogram method}

Not only the temperature ($T$) and chemical potential ($\mu$) 
but also the quark masses are important to understand 
the properties of QCD phase transition.
In fact, the structure of the phase boundary in $T-\mu$ phase diagram is rather
sensitive to the value of the strange quark mass.
Recent lattice QCD simulations suggest that, for physical quark masses,
finite $T$ transition is crossover at zero $\mu$, 
while it becomes first order for sufficiently large $\mu$. 
Identifying the critical point separating crossover
and first order is one of the most challenging topics
in lattice QCD simulations and in heavy-ion experiments.
Probability distribution function or the histogram of the 
order parameter provides us with an important clue to identify such
point in numerical simulations:
In the case of the first order transition, different phases coexist 
at the transition point, so that the probability
distribution function has multiple peaks. 
On the other hand, in the case of crossover, 
such phenomena does not take place.
Therefore, the nature of the transition can be identified 
through the shape of the distribution function.

In this paper, we study the boundary of the first order transition region in QCD in the case when quarks are all heavy. We determine the boundary as function of the chemical potential $\mu$ by measuring histograms.
Although this boundary in the heavy quark region is irrelevant to the boundary near the physical point, 
this provides us with a good testing and developing ground for the method, because the computational burden is much lighter.

Selecting a physical quantity $X$, we calculate the probability distribution function defined by 
\begin{eqnarray}
w(X, \beta, \kappa_f, \mu_f) 
&=& \int {\cal D} U {\cal D} \psi {\cal D} \bar{\psi} \ \delta(X- \hat{X}) \ e^{- S_q - S_g} 
= \int {\cal D} U \ \delta(X- \hat{X}) \ 
e^{-S_g}\ \prod_{f=1}^{N_{\rm f}} \det M(\kappa_f, \mu_f) \nonumber \\
&=& w(X, \beta, 0, 0) 
\left\langle \prod_{f=1}^{N_{\rm f}} \frac{\det M(\kappa_f, \mu_f)}{\det M(0, 0)}
\right\rangle_{(X \ {\rm fixed}; \beta)} ,
\label{eq:dist}
\end{eqnarray}
where $S_g$, $S_q$ and $\det M$ are the gauge action, the quark action and the quark determinant, respectively. 
$\kappa_f$ is the hopping parameter for the $f^{\rm th}$ flavor quark mass.
$\beta=6/g^2$ is the gauge coupling, and 
$N_{\rm f}$ is the number of  flavors.
$\langle \cdots \rangle_{(X \ {\rm fixed}; \beta)} \equiv 
\langle \cdots \delta (X- \hat{X})\rangle_{\beta} / 
\langle \delta(X-\hat{X}) \rangle_{\beta}$ 
means the expectation value measured with fixing the operator $\hat{X}$ at $\beta$ in quenched simulations, $\kappa_f=\mu_f=0$.
The expectation value in the right hand side is the ratio of 
$w(X,\beta, \kappa_f, \mu_f)$ and $w(X,\beta, 0, 0)$. 
However, the calculation of $\det M$ is usually difficult. We perform the hopping parameter expansion and compute the quark determinant in the leading order of the expansion \cite{whot11},
\begin{equation}
\frac{\det M(\kappa, \mu)}{\det M(0,0)} 
= \exp \left[ 
288N_{\rm site} \kappa^4 \hat{P} 
+3N_{\rm s}^32^{N_{\rm t}+2} \kappa^{N_{\rm t}}
\left\{\cosh \left(\frac{\mu}{T}\right) \hat\Omega_{\rm R}
+i\sinh \left( \frac{\mu}{T} \right) \hat\Omega_{\rm I}\right\} 
+ \cdots \right]
\label{eq:detM}
\end{equation}
for the standard Wilson quark action, where $\det M(0,0) = 1$.
The number of sites is $N_{\rm site}=N_s^3 \times N_t$.
The quark determinant is simply given by the average plaquette operator $\hat{P}$ and the real and imaginary parts of the Polyakov loop operator, $\hat{\Omega} = \hat{\Omega}_{\rm R} +i \hat{\Omega}_{\rm I}$.
Because the critical $\kappa$ is very small, at least for $N_t=4$, this approximation can be justified for the determination of the critical $\kappa$.

In this calculation, it is essential to perform simulations at several simulation points and to combine these data by the multi-point reweighting method \cite{FS1988}. 
Since values of the most observables distribute in a narrow range during one Monte-Carlo simulation, it is difficult to investigate the shape of the distribution in a wide range.
We thus combine several simulations.
The expectation value of a operator $\hat{X}$ at $\beta$ is computed by simulations at $\beta_i$ with 
the number of configuration $N_i$ for $i=1, \cdots, N_{\rm SP}$ using the following equation for the case of the plaquette action and degenerate $N_{\rm f}$-flavor,
\begin{eqnarray}
\left\langle \hat{X} \right\rangle_{(\beta, \kappa)}
= \frac{1}{\cal Z} \int {\cal D} U \ \hat{X} 
e^{-S_g} (\det M(\kappa, \mu))^{N_{\rm f}}
=
\frac{ \left\langle \hat{X} \hat{G}(\hat{P}) 
[\det M(\kappa, \mu) / \det M(0, 0)]^{N_{\rm f}} \right\rangle_{\rm all}
}{ \left\langle \hat{G}(\hat{P}) [\det M(\kappa, \mu) 
/ \det M(0, 0)]^{N_{\rm f}} \right\rangle_{\rm all}},
\label{eq:evmultb}
\end{eqnarray}
where the weight factor $\hat{G}(\hat{P})$ is
\begin{eqnarray}
\hat{G}(\hat{P})=\frac{e^{6 N_{\rm site} \beta \hat{P}}}{
\sum_{i=1}^{N_{\rm sp}} N_i e^{ {6 N_{\rm site} \beta_i \hat{P}}} 
{\cal Z}^{-1} (\beta_i)} ,
\end{eqnarray}
and $\left\langle \cdots \right\rangle_{\rm all}$ means the average over 
all configurations generated at all $\beta_i$ with $\kappa=\mu=0$.
The partition functions ${\cal Z} (\beta_i)$ are parameters in this method and 
are determined by solving a consistency condition:
${\cal Z} (\beta_i) 
\approx \sum_{\rm \{ all \ conf. \}} \hat{G}(\hat{P})$ ,
numerically for each $i=1, \cdots, N_{\rm SP}$, except for an overall normalization constant. 
$\sum_{\rm \{all \ conf. \}}$ means the sum of configurations at all $\beta_i$.
(See the appendix A in Ref.~\cite{ejiri08} for details.)
Note that this method enables us to change $\kappa$ and $\beta$ continuously.

\section{Polyakov loop distribution function at zero density}

\begin{figure}[t] 
\begin{minipage}{7cm}
\vspace{-4mm}
\centerline{
\includegraphics[width=85mm]{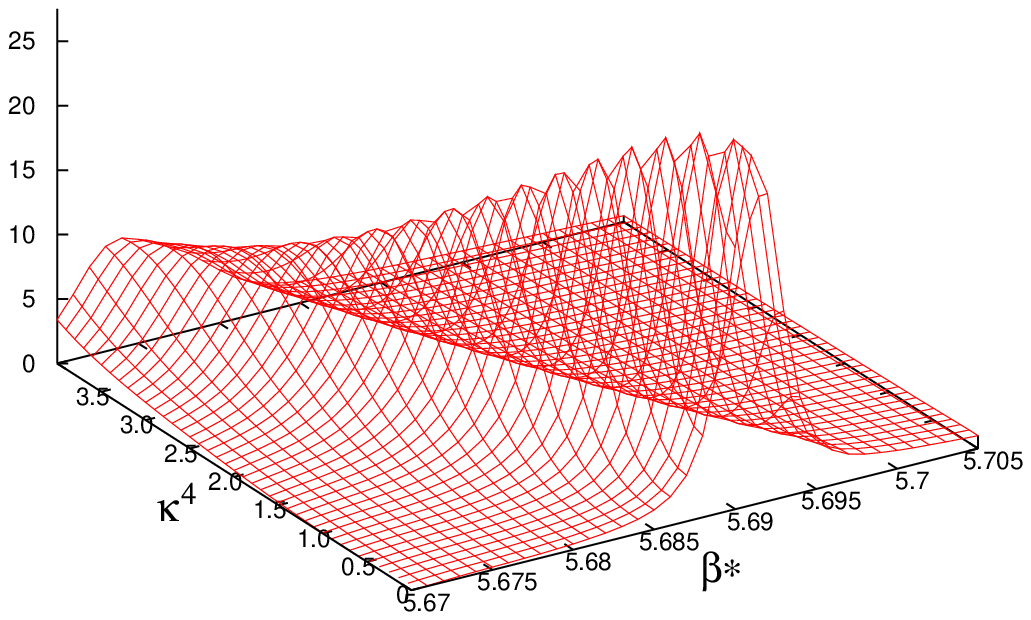} 
}
\caption{Polyakov loop susceptibility as a function of $\kappa^4$ and $\beta^* = \beta+48N_{\rm f} \kappa^4$ for $N_{\rm f}=2$.}
\label{fig:plsus}
\end{minipage}
\hspace{0.8cm}
\begin{minipage}{7cm}
\centerline{
\includegraphics[width=6.7cm]{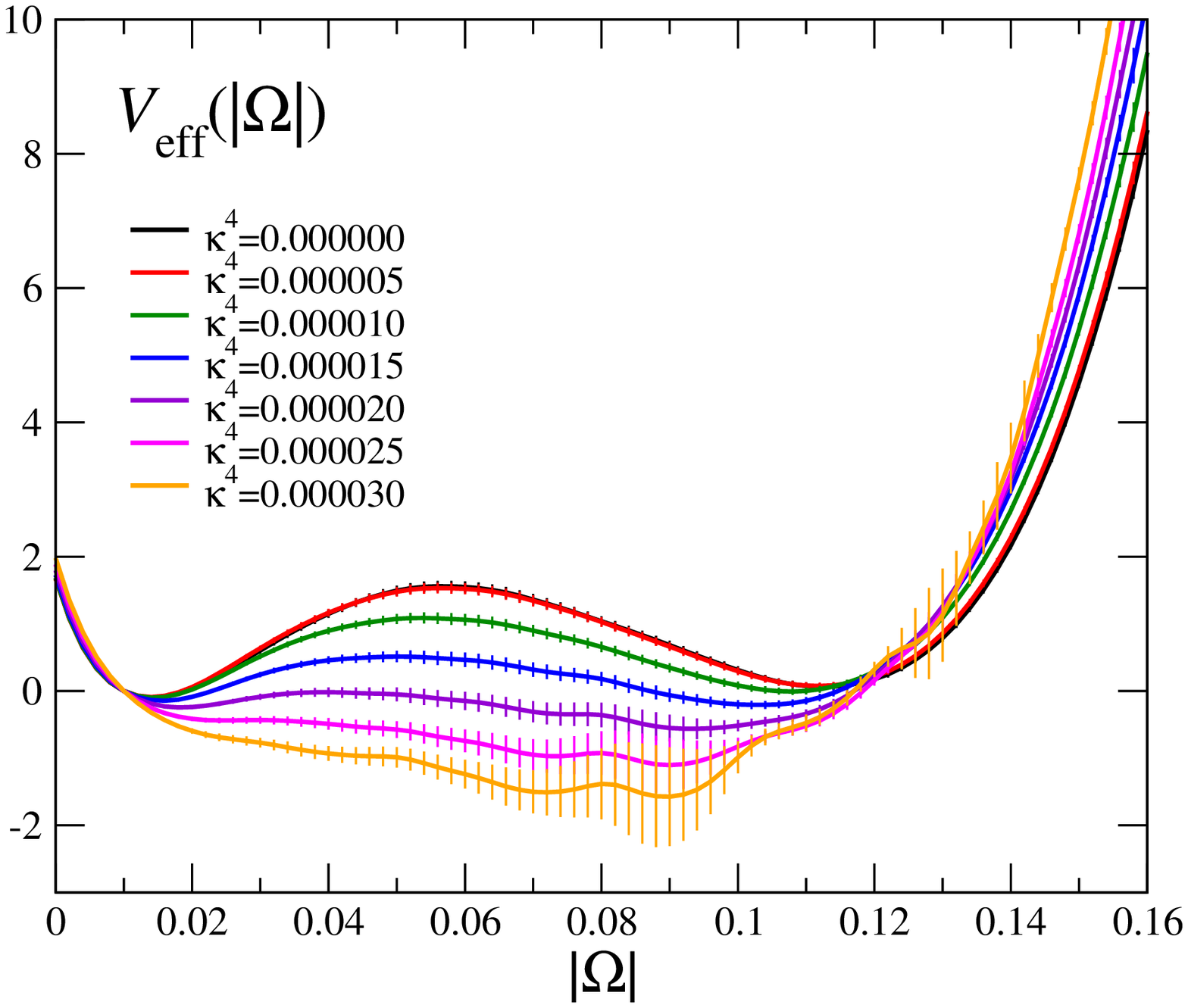}
}
\caption{Distribution function of the absolute value of the Polyakov loop at the transition point.}
\label{fig:oabhist}
\end{minipage}
\end{figure}

The most important observable near the transition point in the heavy quark region is the Polyakov loop, which is the order parameter of the deconfinement transition.
We analyze the data obtained at 5 simulation points, $\beta =5.68$ -- $5.70$, in the quenched simulations with the plaquette gauge action on a $24^3 \times 4$ lattice \cite{whot11}.
Figure \ref{fig:plsus} is the result of the Polyakov loop susceptibility, 
$\chi_{\Omega} =N_{s}^3 \langle (\Omega - \langle \Omega \rangle)^2 \rangle$,
as a function of $\kappa^{N_t}$ and the effective $\beta$ defined as $\beta^* = \beta +48N_{\rm f} \kappa^4$ with $N_t=4$, computed at $\mu=0$ using Eq.~(\ref{eq:evmultb}).
Because the plaquette action is $S_g=-6N_{\rm site} \beta \hat{P}$, the plaquette term in Eq.~(\ref{eq:detM}) can be absorbed into the gauge action by defining $\beta^*$ and the analysis becomes simpler.
Owing to the multi-point reweighting method, the susceptibility can be calculated in a wide range of $\beta$ and $\kappa$. We define the transition point as the peak position of $\chi_{\Omega}$.

We then measure the distribution function of the absolute value of the Polyakov loop at the transition point for the case of 2-flavor QCD at $\mu=0$. 
Expectation values with fixing the value of the Polyakov loop are computed using the delta function approximated by a Gaussian function,
$\delta(x) \approx \exp[-(x/\Delta)^2]/(\Delta \sqrt{\pi}) $, 
where $\Delta=0.005$ is adopted consulting the resolution and the statistical error. 
We plot the effective potential,  
$V_{\rm eff}(|\Omega|) = - \ln w(|\Omega|)$,
for several values of $\kappa^4$ in Fig.~\ref{fig:oabhist}. $\beta$ is adjusted to the peak position of $\chi_{\Omega}$ at each $\kappa$.  
The value of $V_{\rm eff}(|\Omega|)$ is normalized at $|\Omega|=0.01$.
This figure shows that the shape of $V_{\rm eff}(|\Omega|)$ is double-well type at $\kappa^4=0$, indicating the first order transition,
and the shape changes gradually as increasing $\kappa$.
It becomes single-well around $\kappa^4 \sim 0.00002$, suggesting the first order transition changes to crossover. 
The critical value of $\kappa$ has been determined by measuring the distribution function of the average plaquette in Ref.~\cite{whot11} with the same configurations.
The result is $\kappa_{\rm cp}=0.0658(3)(^{+4}_{-11})$ for $N_{\rm f}=2$.
Hence, the results of $\kappa_{\rm cp}$ from the plaquette and Polyakov loop effective potentials are consistent with each other.

\begin{figure}[tb]
\vspace{-9mm}
\centerline{
\includegraphics[width=59mm]{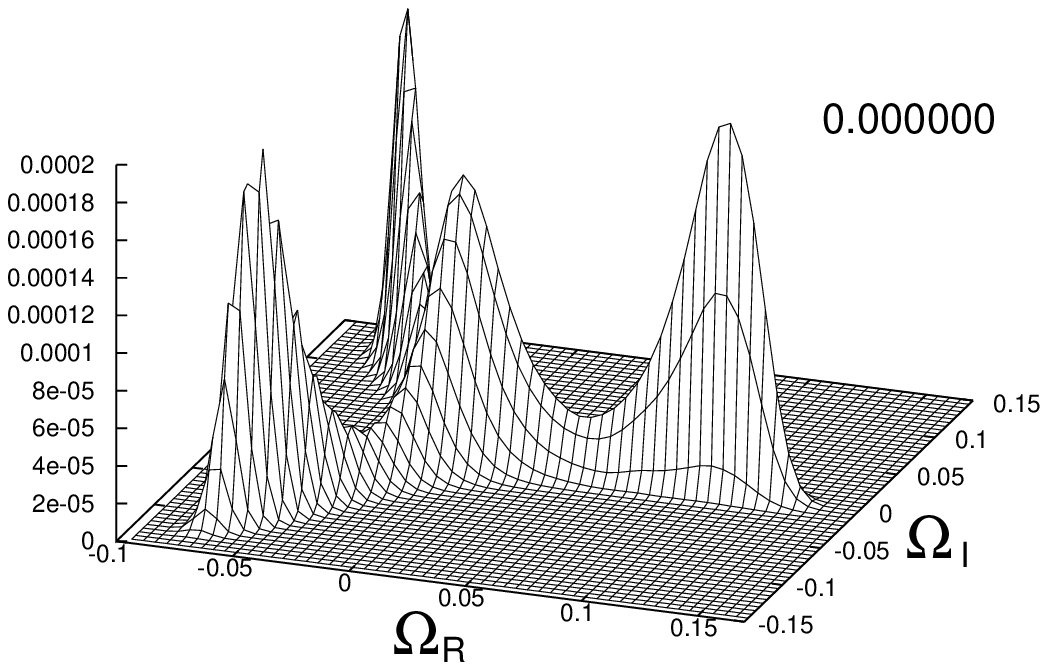}
\hspace{-11mm}
\includegraphics[width=59mm]{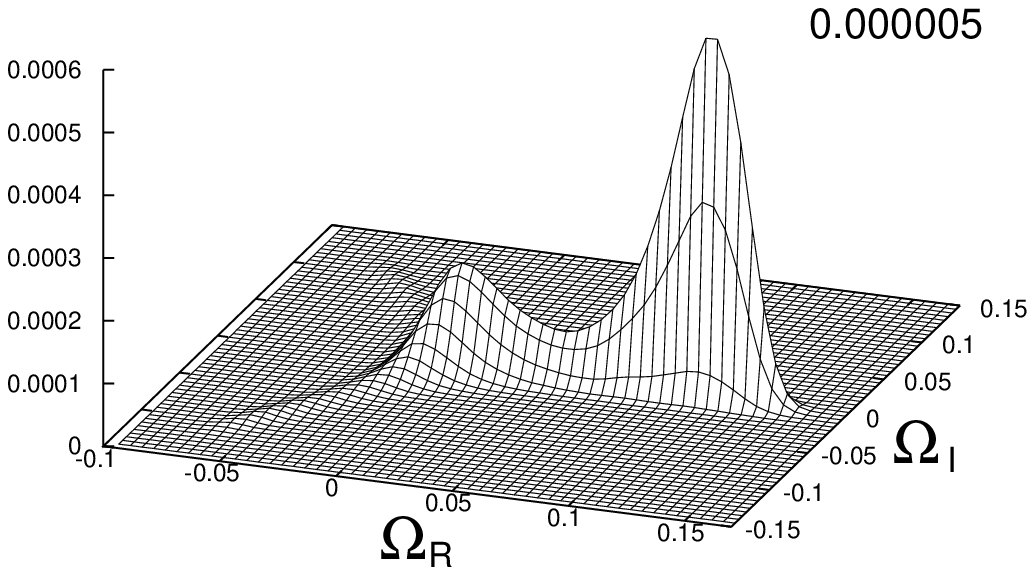}
\hspace{-11mm}
\includegraphics[width=59mm]{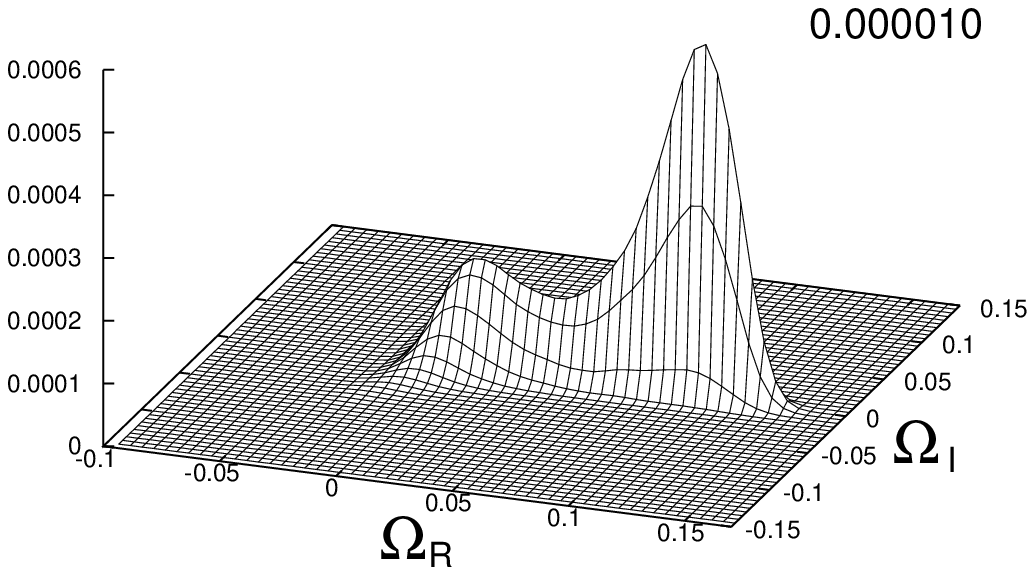}
}
\vspace{-12mm}
\centerline{
\includegraphics[width=59mm]{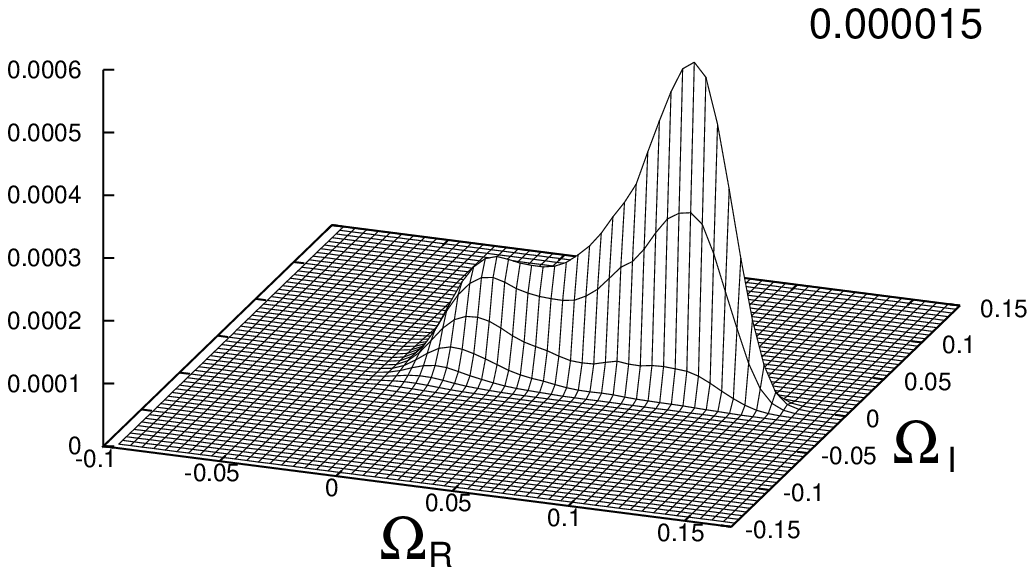}
\hspace{-11mm}
\includegraphics[width=59mm]{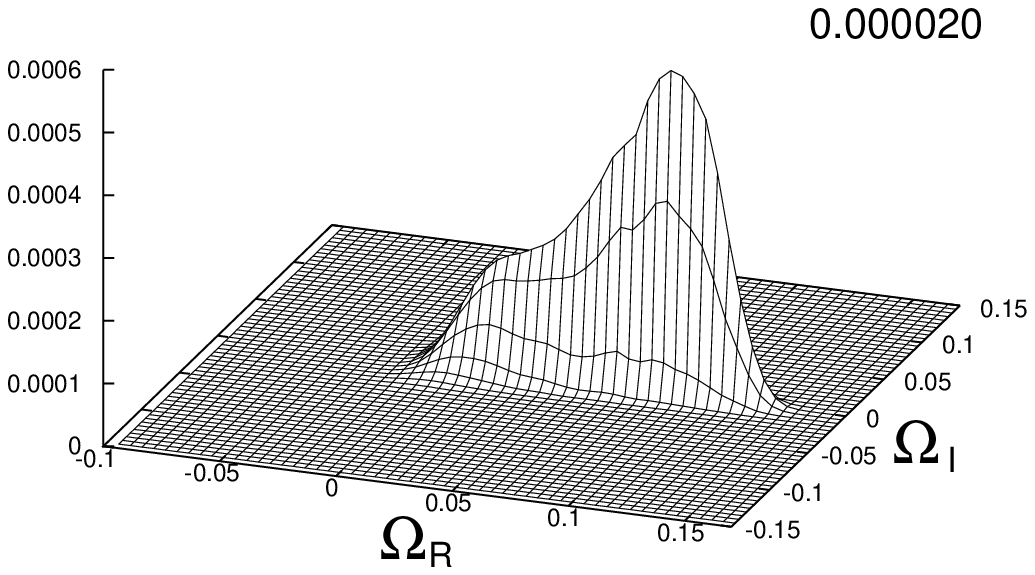}
\hspace{-11mm}
\includegraphics[width=59mm]{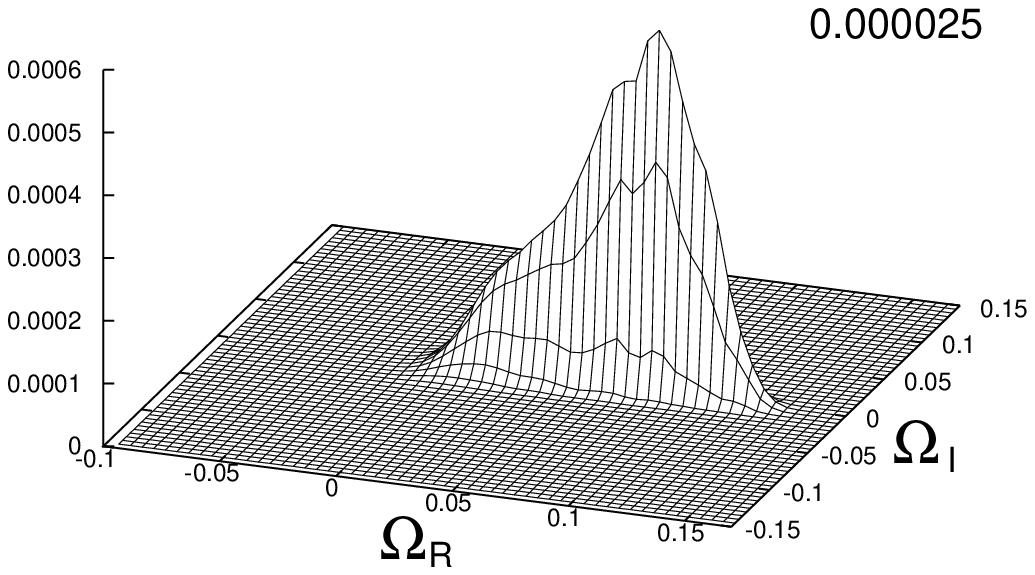}
}
\vspace{-2mm}
\caption{$\kappa$-dependence of the probability distribution of the Polyakov loop at the transition point in the complex plane for $N_{\rm f}=2$.
The value of $\kappa^4$ is shown in the upper right of the figure.
}
\label{fig:compol}
\end{figure}

We moreover calculate the distribution function of the complex Polyakov loop in the complex plane $(\Omega_{\rm R}, \Omega_{\rm I})$ at the phase transition point, which is shown in Fig.~\ref{fig:compol} for 2-flavor QCD. 
The well-known Z(3) symmetric 4 peak structure is observed at $\kappa=0$, and the 2 peaks in the negative $\Omega_{\rm R}$ region become smaller as increasing $\kappa$. 
Then, the remaining 2 peaks are getting closer with $\kappa$, and the distribution becomes a single peak around $\kappa=0.00002$.
These figures illustrate how the Z(3) symmetric quenched QCD changes to full QCD.

\section{Complex phase and distribution function at finite density}

The histogram method is powerful in particular with the presence of the chemical potential $\mu$.
Direct simulations by the Monte Carlo method cannot be performed at finite chemical potential because the quark determinant is complex. 
An approach to simulate finite density QCD is to combine the reweighting method and simulations with the complex phase of the quark determinant suppressed, which are called phase-quenched simulations.
The distribution function for the real part of Polyakov loop, $\Omega_{\rm R}$, is a good example to explain the contribution from the complex phase and the phase-quenched part in the distribution function at finite density.

We calculate the distribution function in heavy quark QCD for the degenerate $N_{\rm f}=2$ standard Wilson case.
Using the hopping parameter expansion, the distribution function can be factorized into the phase factor and the phase-quenched part. 
\begin{eqnarray}
w(\Omega_{\rm R}, \beta, \kappa, \mu) 
&=& \int {\cal D} U \ \delta(\Omega_{\rm R} - \hat{\Omega}_{\rm R}) \ 
e^{6N_{\rm site} \beta \hat{P}} \ (\det M(\kappa, \mu))^{N_{\rm f}} \nonumber \\
&& \hspace{-25mm} = w(\Omega_{\rm R}, \beta, 0, 0) 
\left\langle  e^{288N_{\rm site} N_{\rm f} \kappa^4 \hat{P}} 
\exp \left[ 3 N_s^3 2^{N_t+2} N_{\rm f}
\kappa^{N_t} \left\{ \cosh \left( \frac{\mu}{T} \right) \hat{\Omega}_{\rm R}
+i\sinh \left( \frac{\mu}{T} \right) \hat{\Omega}_{\rm I} \right\} \right]
\right\rangle_{(\Omega_{\rm R}; \beta, \kappa)} \nonumber \\
&& \hspace{-25mm} = w(\Omega_{\rm R}, \beta^*, 0, 0) 
\exp \left[ 3 N_s^3 2^{N_t+2}  N_{\rm f}
\kappa^{N_t} \cosh \left( \frac{\mu}{T} \right) \Omega_{\rm R} \right] 
\left\langle e^{i \hat{\theta}} \right\rangle_{(\Omega_{\rm R}; \beta^*, 0)} ,
\label{eq:plhvdist}
\end{eqnarray}
where $\hat{\theta}$ is the phase of the quark determinant: 
\begin{eqnarray}
\hat{\theta} =3 N_s^3 2^{N_t+2} N_{\rm f} \kappa^{N_t} 
\sinh(\mu/T) \ \hat{\Omega_{\rm I}},
\end{eqnarray}
and the part in front of the phase average is the distribution function in the phase-quenched theory.
The plaquette term is absorbed by $S_g$ shifting $\beta$ to 
$\beta^* = \beta +48 N_{\rm f} \kappa^4$.
$\langle \cdots \rangle_{(\Omega_{\rm R}; \beta, \kappa)}$ means the expectation value at $(\beta, \kappa)$ with fixing $\Omega_{\rm R}$. 

\begin{figure}[t] 
\begin{minipage}{7cm}
\centerline{
\includegraphics[width=67mm]{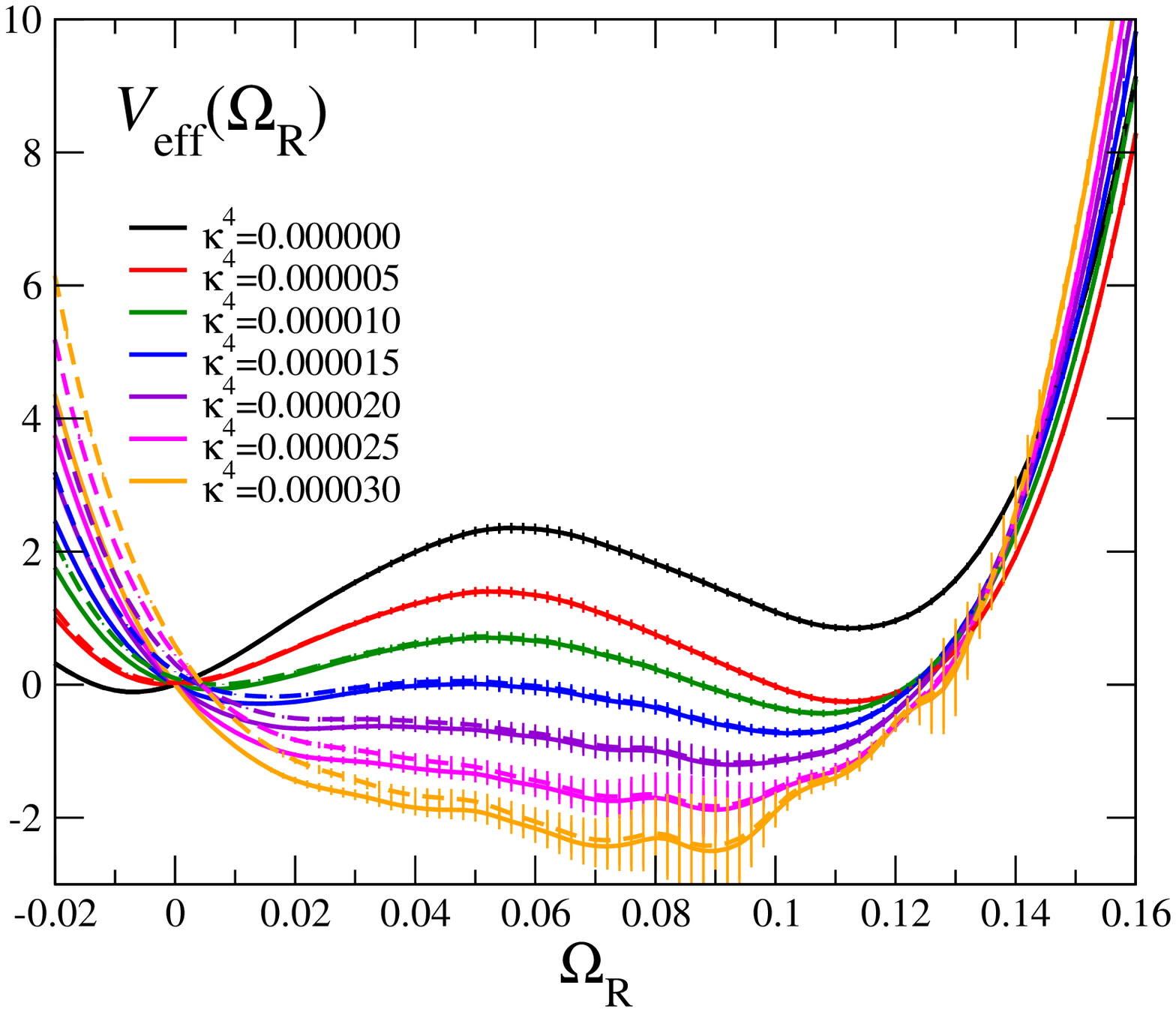}
}
\caption{The solid lines are $V_{\rm eff} (\Omega_{\rm R})$ at $\mu=0$ for each $\kappa^4$. 
$V_{\rm eff} (\Omega_{\rm R})$ at finite $\kappa^4 \cosh (\mu/T)$ is between the solid line 
and the dashed line.}
\label{fig:orhist}
\end{minipage}
\hspace{0.8cm}
\begin{minipage}{7cm}
\centerline{
\includegraphics[width=67mm]{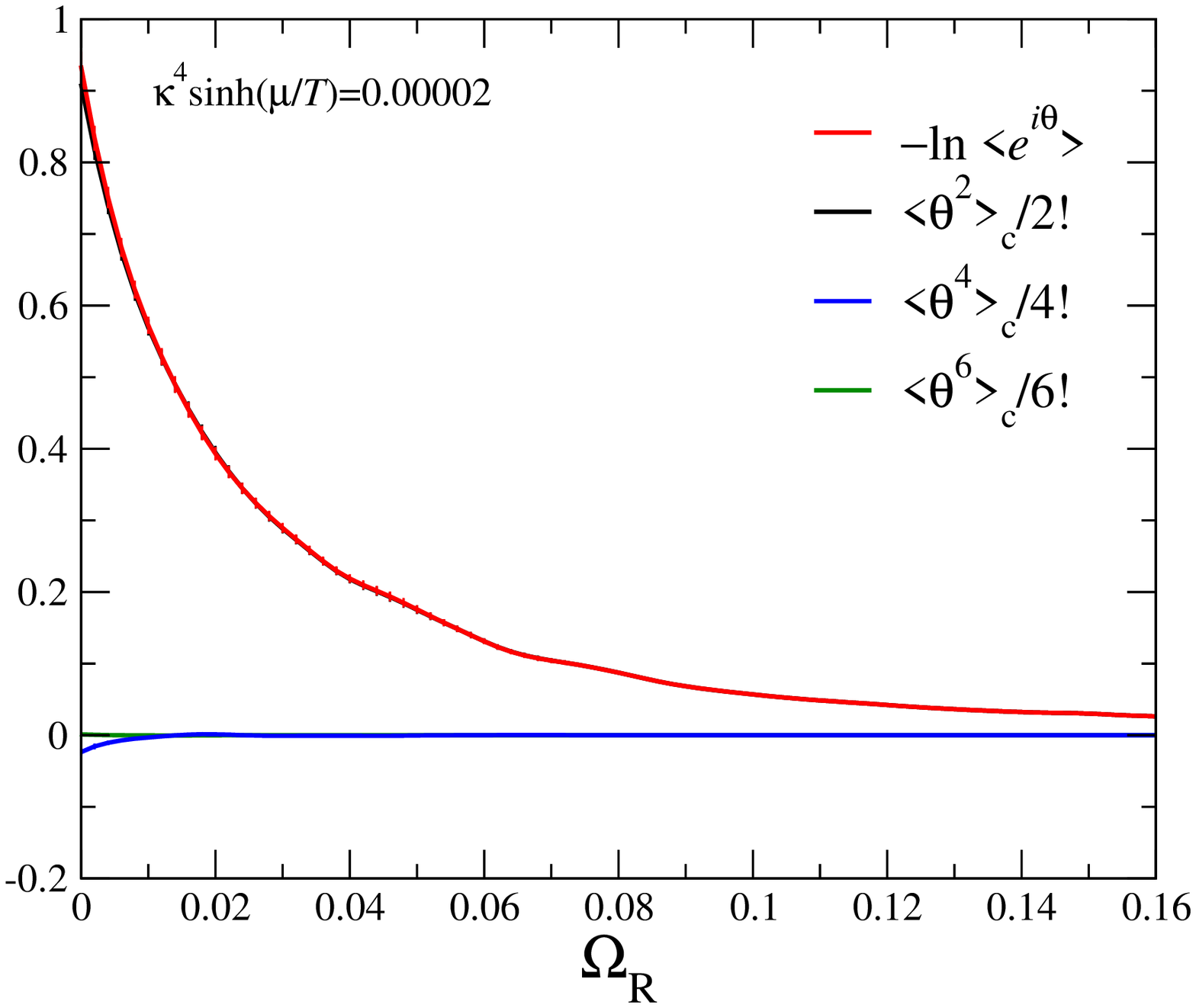} 
}
\caption{The average of the complex phase factor and the $2^{\rm nd}$, $4^{\rm th}$ and $6^{\rm th}$ order cumulants 
calculated with fixed $\Omega_{\rm R}$ at $\kappa^4 \sinh (\mu/T) \approx 0.00002$.}
\label{fig:phase}
\end{minipage}
\end{figure}

The phase-quenched part of $w(\Omega_{\rm R}, \beta, \kappa, \mu)$
can be obtained from that at $\mu=0$ simply by replacing $\kappa^{N_t}$ by $\kappa^{N_t} \cosh(\mu/T)$, 
since the distribution function at $\mu=0$ is given by 
$ w(\Omega_{\rm R}, \beta^*, 0, 0)$ $\times \exp [ 3 N_s^3 2^{N_t+2}  \kappa^{N_t} \Omega_{\rm R} ] $.
Therefore, the critical value $\kappa_{\rm cp} (\mu)$ in the phase-quenched theory is given by 
$\kappa_{\rm cp}^{N_t}(0) = \kappa_{\rm cp}^{N_t}(\mu) \cosh(\mu/T)$.
Moreover, adopting $\kappa^{N_t} \cosh(\mu/T)$ as basic parameter to investigate the critical point, 
the magnitude of the phase is limited for each $\kappa^{N_t} \cosh(\mu/T)$ 
because $\kappa^{N_t} \sinh(\mu/T)$ in $\hat{\theta}$ is always smaller than $\kappa^{N_t} \cosh(\mu/T)$.
The effective potential of $\Omega_{\rm R}$, $V_{\rm eff} (\Omega_{\rm R}) = - \ln w (\Omega_{\rm R})$, at the transition point for $\mu=0$ is plotted in Fig.~\ref{fig:orhist} by the solid lines for each $\kappa^{N_t}$, and is equal to the phase-quenched distribution function for each $\kappa^{N_t} \cosh(\mu/T)$.
This $V_{\rm eff}$ is normalized at $\Omega_{\rm R} =0$.

Next, we calculate the phase factor, 
$\langle e^{i \hat{\theta}} \rangle_{(\Omega_{\rm R}; \beta^*, 0)}$.
If $e^{i \hat{\theta}}$ changes its sign frequently, the statistical error becomes larger than the expectation value, causing the sign problem.
To avoid the sign problem, we evaluate the phase factor by the cumulant expansion \cite{ejiri07,whot10}: 
\begin{eqnarray}
\left\langle e^{i \hat\theta} \right\rangle_{(\Omega_{\rm R}; \beta^*, 0)} \; = \;
\exp \left[i \langle \hat\theta \rangle_c
- \frac{\langle \hat\theta^2 \rangle_c}{2} 
- \frac{i \langle \hat\theta^3 \rangle_c}{3!} 
+ \frac{\langle \hat\theta^4 \rangle_c }{4!} 
+ \frac{i \langle \hat\theta^5 \rangle_c }{5!} 
- \frac{\langle \hat\theta^6 \rangle_c}{6!} + \cdots \right],
\label{eq:cum}
\end{eqnarray}
where $\langle \hat\theta^n \rangle_c$ is the $n^{\rm th}$ order cumulant:
$
\langle \hat\theta^2 \rangle_c 
= \langle \hat\theta^2 \rangle_{(\Omega_{\rm R}; \beta^*, 0)} 
$, 
$
\langle \hat\theta^4 \rangle_c 
= \langle \hat\theta^4 \rangle_{(\Omega_{\rm R}; \beta^*, 0)}
-3 \langle \hat\theta^2 \rangle_{(\Omega_{\rm R}; \beta^*, 0)}^2
$, 
$
\langle \hat\theta^6 \rangle_c 
= \langle \hat\theta^6 \rangle_{(\Omega_{\rm R}; \beta^*, 0)}
-15 \langle \hat\theta^4 \rangle_{(\Omega_{\rm R}; \beta^*, 0)} 
\langle \hat\theta^2 \rangle_{(\Omega_{\rm R}; \beta^*, 0)} 
+30 \langle \hat\theta^2 \rangle_{(\Omega_{\rm R}; \beta^*, 0)}^3 
$, $\cdots$.
The key point of this method is that $\langle \hat\theta^n \rangle_c =0$ for any odd $n$
due to the symmetry under $\hat\theta \rightarrow -\hat\theta$, 
and thus the complex phase can be omitted from this equation.
This implies that $\langle e^{i \hat\theta} \rangle$ is guaranteed to be real and positive and the sign problem is resolved once the cumulant expansion converges.
Another important point is that $\hat{\theta}$ is given by the average of the Polyakov loop.
When the correlation length is finite, the phase can be written as a summation of local 
contributions $\hat{\theta}= \sum_x \hat{\theta}_x$ with almost independent $\hat{\theta}_x$.
The phase average is then
\begin{eqnarray}
\left\langle e^{i\hat\theta} \right\rangle 
\approx \prod_x \left\langle e^{i\hat\theta_x} \right\rangle 
= \exp \left( \sum_x \sum_n \frac{i^n}{n!} 
\left\langle \hat\theta_x^n \right\rangle_c \right).
\end{eqnarray}
This suggests that all cumulants $\langle \hat\theta^n \rangle_c 
\approx \sum_x \langle \hat\theta_x^n \rangle_c$ 
increase in proportion to the volume as the volume increases.
Only for such a case, the effective potential $V_{\rm eff}$ 
can be well-defined though 
the phase-quenched effective potential $V_0$ with 
$V_{\rm eff} = V_0 - \ln \langle e^{i \hat{\theta}} \rangle = V_0 - \sum_n i^{n} \langle \hat{\theta}^{n} \rangle_c/n!$ in the large volume limit, 
since $V_{\rm eff}$ and $V_0$ are both in proportion to the volume. 

We plot the results of $\langle \hat\theta^n \rangle_c /n!$ in Fig.~\ref{fig:phase} for $\kappa^{N_t} \sinh (\mu/T)=0.00002$ and $\beta^* =5.69$.
The black, blue and green lines are the results for $n=2, 4$ and 6, respectively.
The fourth and sixth order cumulants are very small in comparison to the second order for this $\kappa$. 
The red line is $- \ln \langle e^{i \hat{\theta}} \rangle_{(\Omega_{\rm R}; \beta^*, 0)}$, 
which is almost indistinguishable from the second order cumulant.
The contribution from the fourth and sixth orders becomes visible at small $\Omega_{\rm R}$ as $\kappa^{N_t} \sinh (\mu/T)$ increases. 
However, for the determination of the critical point, 
$\kappa_{\rm cp}^{N_t} \cosh (\mu/T) \approx 0.00002$ in the phase-quenched theory, the region at 
$\kappa^{N_t} \sinh (\mu/T) < 0.00002$ is important because $\cosh (\mu/T) > \sinh (\mu/T)$. 
This figure thus indicates that the phase average is well-approximated 
by the second order cumulant around the critical $\kappa$. 
The results of the effective potentials including the effect from the phase factor are shown by the dashed lines in Fig.~\ref{fig:orhist} for each $\kappa^{N_t} \cosh (\mu/T)$. 
In this figure, the phase factor is estimated by the second order cumulant at $\mu/T= \infty$, i.e. $\sinh (\mu/T)=\cosh (\mu/T)$. 
The results at finite $\mu$ are between the solid and dashed lines.
We find that the contribution from the phase, 
$-\ln \langle e^{i \hat{\theta}} \rangle$, 
is quite small except at small $\Omega_{\rm R}$ and the phase factor does not affect $V_{\rm eff}$ in the region of $\Omega_{\rm R}$ relevant for the determination of the critical point. 
This means that the contribution from the complex phase to the location of the critical point is quite small on our $24^3 \times 4$ lattice.

\begin{figure}[tb]
\centerline{
\includegraphics[width=76mm]{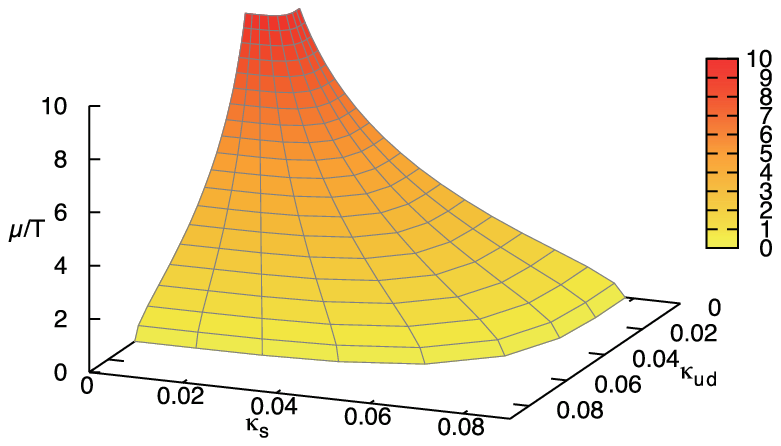}
\includegraphics[width=76mm]{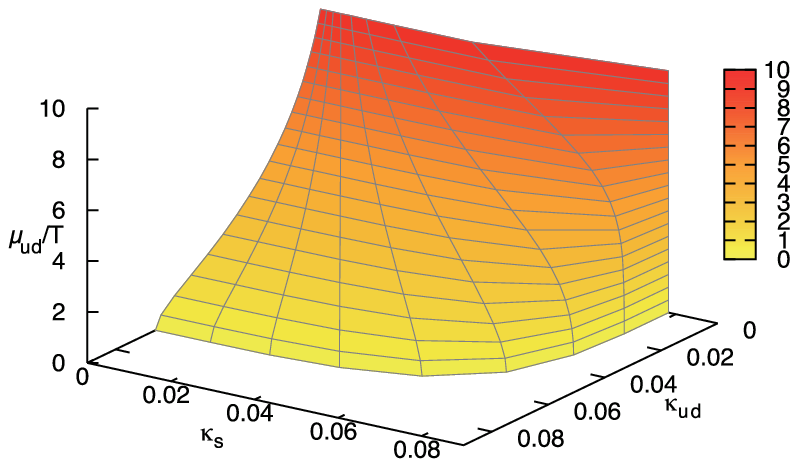}
}
\caption{Critical surface separating the first order transition and crossover regions in the heavy-quark region.
 {\em Left}: 
The case $\mu_{u} = \mu_{d} = \mu_{s} \equiv \mu$.
 {\em Right}: 
The case that may be realized in heavy ion collisions: $\mu_{u} = \mu_{d} \equiv \mu_{ud}$ and $\mu_{s} = 0$.
}
\label{fig:crtsur}
\end{figure}

Neglecting the effect of the phase factor,
it is easy to determine the critical point for the $N_{\rm f}=2+1$ case
because the difference from the $N_{\rm f}=2$ case is just to replace $2\kappa^{N_t}$ by $2 \kappa_{\rm ud}^{N_t} + \kappa_{\rm s}^{N_t}$. 
We thus find that the critical $(\kappa_{\rm ud}, \kappa_{\rm s})$ is given by 
\begin{equation}
2 \kappa_{\rm ud}^{N_t}(\mu) \cosh(\mu_{\rm ud}/T) 
+ \kappa_{\rm s}^{N_t}(\mu) \cosh(\mu_{\rm s}/T) 
= 2 [\kappa_{\rm cp}^{N_{\rm f}=2}(0) ]^{N_t} ,
\end{equation}
where $\kappa_{\rm cp}^{N_{\rm f}=2}(0)=0.0658(3)(^{+4}_{-11})$ for $N_t=4$ 
\cite{whot11}.
The critical lines in the $\kappa$ plane for up, down and strange are drown 
in Fig.~\ref{fig:crtsur} 
for the cases of $\mu_{\rm ud}/T=\mu_{\rm s}/T=0$ -- $10$ (left) and 
$\mu_{\rm ud}/T=0$ -- $10$ and $\mu_{\rm s}/T=0$ (right).

\section{Summary}

We have studied the phase structure of QCD at nonzero chemical potential $\mu$ in the heavy quark region, highlighting the properties of the probability distribution function of the Polyakov loop $\Omega$.
The shape of the effective potential defined by the distribution function changes with the quark mass and the chemical potential. 
The multi-point reweighting technique enables us to obtain the distribution function in a wide range of the coupling parameters.
We have shown that the effective potential provides us with an intuitive and powerful way to investigate the fate of first order phase transitions.
Through the shape of the potential, the critical surface where the first order deconfining transition in the heavy quark limit terminates is determined for the 2+1-flavor case. 
The effect from the complex phase of the quark determinant has been estimated explicitly, and is found to be quite small around the critical point for any chemical potential in the heavy quark region.
On the other hand, the effect from the complex phase must be important in the light quark region. 
An attempt to study finite density QCD at light quark masses by combining phase-quenched simulations and the reweighting technique is reported in \cite{nakagawa}.

\vspace{5mm}

This work is supported in part by Grants-in-Aid of the Japanese Ministry
of Education, Culture, Sports, Science and Technology 
(Nos.~20340047,  
21340049 , 
22740168, 
23540295)  
and by the Grant-in-Aid for Scientific Research on Innovative Areas
(Nos.~2004:20105001, 20105003, 2310576). 

\end{document}